\RequirePackage{lineno}
\documentclass[aps,prl,nofootinbib,twocolumn,superscriptaddress,letterpaper,amsmath,amssymb]{revtex4-1}

\newcommand{\jsdbefore}{JSD\textsubscript{before }}
\newcommand{\jsdafter}{JSD\textsubscript{after }}
\newcommand\ph{$\phantom{1}$}
\newcommand\varset{\ensuremath{\{x_n\}}}

\usepackage{graphicx}

\begin{document}

\title[On the Upper Limit of Separability]{On the Upper Limit of Separability}

\author{Nick Carrara}
\author{Jesse Ernst}
\affiliation{Physics Department, University at Albany, SUNY, Albany, NY 12222}
\email{jae@albany.edu, ncarrara@albany.edu}

\begin{abstract}

We propose an approach to rapidly find the upper limit of separability between
datasets that is directly applicable to HEP classification problems. The most
common HEP classification task is to use $n$ values (variables) for an object
(event) to estimate the probability that it is signal vs.\ background. Most
techniques first use known samples to identify differences in how signal and
background events are distributed throughout the $n$-dimensional variable
space, then use those differences to classify events of unknown type.
Qualitatively, the greater the differences, the more effectively one can
classify events of unknown type. We will show that the Mutual Information (MI)
between the $n$-dimensional signal-background mixed distribution and the
answers for the known events, tells us the upper-limit of separation for that
set of $n$ variables. We will then compare that value to the Jensen-Shannon
Divergence between the output distributions from a classifier to test whether
it has extracted all possible information from the input variables. We will
also discuss speed improvements to a standard method for calculating MI.

Our approach will allow one to: a) quickly measure the maximum possible
effectiveness of a large number of potential discriminating variables
independent of any specific classification algorithm, b) identify potential
discriminating variables that are redundant, and c) determine whether a
classification algorithm has achieved the maximum possible separation. We test
these claims first on simple distributions and then on Monte Carlo samples
generated for Supersymmetry and Higgs searches. In all cases, we were able to
a) predict the separation that a classification algorithm would reach, b)
identify variables that carried no additional discriminating power, and c)
identify whether an algorithm had reached the optimum separation. Our code is
publicly available.

\end{abstract}

\maketitle


\begin{section}{Introduction}

From particle physics to finance to medical diagnosis, classification problems
are ubiquitous. In one of the most common HEP tasks, one seeks to place
objects (events) into one of two categories ($\{\theta_m\}$, with $m=+1,-1$
for signal and background respectively) based on a set \varset\ of $n$
measured values (variables) for each event. One often selects the variables
used for classification by choosing those that each show some ability to
discriminate between the two classes, i.e., a variable is used if its
distribution differs between signal and background events. However, the
process of choosing variables, known as feature selection, can be difficult
because several variables that each individually show little difference
between signal and background may, because of correlations with one another,
be useful in classification when used together~\cite{Guyon}--\cite{Vergara}.

Independent of the specific classification method, one typically begins with
samples of both event types (known samples). These allow one to characterize
differences in how signal and background events are distributed throughout the
$n$-dimensional variable space. Once those differences are characterized, one
uses the result to classify unknown events. In neural network methods, for
example, one first uses the known samples to fix the parameters of the network
model, and later uses that model to classify events of unknown type.
Qualitatively, if an event of unknown type lies in a region of variable space
that was populated more heavily with signal events than background events in
the known samples, then the unknown event is more likely to be a signal event.
Thus the fundamental question of signal-background separability is this: In
the $n$-dimensional variable space, how different are the signal and
background distributions? The extreme cases of complete separation and zero
separation lead to perfect distinguishability and no distinguishability
respectively. Any realistic problem will lie between these extremes.

To find a fundamental measure for the separability between datasets we begin
with the following simple idea: When using \varset\ to decide whether an event
is signal or background, how much additional information would the answer
give? Consider, for example, signal and background events where a single
variable is used to discriminate between them. If there were no overlap of the
signal and background distributions for the variable, then its value is as
useful as the answer itself for classifying an event, and thus the answer
would provide no additional information. At the other extreme, if the two
classes of events overlap completely in the variable, then the answer would
provide all the discriminating information and so its addition gives the
maximum increase in information about the event type. One can easily extend
this notion to the realistic case of partially overlapping distributions in a
variable space of $n$-dimensions. Here, one would ask: If one added the answer
to the n discriminating variables, how much information about the event type
would one gain, or inversely, to what extent is the answer
redundant?~\cite{nj_maxent_poster}

Concepts from information theory will allow us to quantify
this. We will first review the concepts of Shannon
Entropy, Cross Entropy, and the Kullback-Leibler Divergence, then discuss the
Jensen-Shannon Divergence and its equivalence to mutual information (MI). We
will show that JSD/MI is invariant under transformations from $n$-dimensions
to 1-dimension as long as the relative signal-background density is maintained
in the transformation. The Neymann-Pearson lemma will allow us to assert that
any optimum classification algorithm will have this property. We will then use
JSD and MI to describe a practical method for calculating a numerical limit on
the separability of signal-background datasets and show the benefits of
comparing it to the performance of a classifier.

We will first test these techniques on simple models with Gaussian
distributions and then apply them to datasets from simulated particle physics
data. Our software for computing separability limits, along with documentation
to help users quickly calculate limits for their own datasets, is publicly
available at \url{https://github.com/albanyhep/JSDML}.

\end{section}

\begin{section}{Shannon Entropy, Kullback-Leibler Divergence, and
Jensen-Shannon Divergence}

If one draws a list of values from a discrete probability distribution and
then attempts to encode it efficiently, the Shannon Entropy~\cite{Shannon},
$H$, is the minimum number of bits per entry that will be needed for an
average list. For a discrete distribution $Q$ with values $q_1, q_2 \dotsc
q_n$, $H(Q)=-\sum_{i=1}^{n} P(q_i)\log_b P(q_i)$, where $P(q_i)$ is the
probability for $q_i$. (We will use log base 2 throughout, which gives $H$
units of bits). To achieve the encoding minimum given by $H$, one needs to
assign values wisely in the encoding scheme. Values corresponding to large
bins in $Q$ will appear more often in a typical list and so should be assigned
to small values (fewer bits) for encoding. Conversely, values corresponding to
small bins in Q will appear infrequently and so should be assigned to larger
values (more bits) for encoding. Unsurprisingly, $H$ is a maximum for uniform
distributions ($H=\log_2(n)$ for $n$ bins, which reduces to zero, as expected,
for a single-bin distribution) and decreases as the distribution becomes less
uniform. The cross entropy $H(P,Q)$ of two distributions $P$ and $Q$ is a
closely related quantity. If one designs the most efficient encoding scheme
for lists drawn from $Q$ but then uses it to encode lists drawn from $P$, the
cross entropy $H(P,Q) = -\sum_{i=1}^{n} P(p_i)\log_2 P(q_i)$ is the number, on
average, of bits per entry needed for the encoding. Unless $P$ and $Q$ are
identical, then encoding lists drawn from $P$, using the encoding scheme
optimized for $Q$, will be less efficient and so will require, on average,
more bits per entry. The number of extra bits per entry, on average, is
$H(P,Q) - H(P)$ and is known as the Kullback-Leibler
divergence~\cite{Kullback_a,Kullback_b} ($D_{kl}(P||Q)$) or the relative
entropy of $P$ with respect to $Q$. Relevant to our work, it is useful to note
that because $H(P)$, $H(P,Q)$, and $D_{kl}$ are all simply sums over bins, the
dimensionality of the distribution is irrelevant. A distribution of $n$ bins
in $d$ dimensions can be rearranged into a distribution of $n$ bins in
1-dimension without changing them.

The Mutual Information (MI) between discrete random variables $X$ and $Y$, is
the extent to which knowing the value of one of them reduces the uncertainty
in the other. In terms of $D_{kl}$, $I(X;Y)=D_{kl}(p(x,y)||p(x)p(y))$ where
$p(x,y)$ is the joint distribution of $x$ and $y$ and $p(x)$, and $p(y)$ are
the marginal distributions. Thus $I(x;y)$ quantifies the relationship between
$x$ and $y$ as the encoding penalty incurred by encoding samples drawn from
$p(x,y)$ under the assumption that $p(x)$ and $p(y)$ are uncorrelated. MI has
often been explored for feature extraction as an alternative to the
correlation coefficient.

Although $D_{kl}$ is a widely used measure of dissimilarity, particularly for
characterizing the change between prior and posterior distributions, a
symmetric measure, Jensen-Shannon Divergence (JSD), is more useful for our
application~\cite{Lin}. The JSD for $P$ and $Q$, normalized to equal numbers
of entries, can be written in terms of $D_{kl}$ as $\text{JSD}(P||Q) =
\frac{1}{2} D_{kl}(P||M) + \frac{1}{2} D_{kl}(Q||M)$, where,
$M=\frac{1}{2}(P+Q)$ is the mixture distribution of $P$ and $Q$. Both JSD and
$D_{kl}$ quantify the difference between $P$ and $Q$, but
$\text{JSD}(P||Q)=\text{JSD}(Q||P)$ while generally $D_{kl}(P||Q) \neq
D_{kl}(Q||P)$. To quantify the difference between signal and background, that
symmetry is necessary. JSD can be understood through its close connection to
MI. The JSD of $P$ and $Q$ is the MI between \varset\ and the answer
$\{\theta\}$ for $M$ ($\text{JSD}(P||Q)=I(M;\theta$)). Qualitatively, it
measures the extent to which knowing \varset\ for an event drawn from $M$
reduces the uncertainty in the answer $\theta$. Or equivalently, the extent to
which knowing the variables makes the answer itself redundant. For completely
overlapping (separate) distributions, the value of the variable(s) for a
single event from $M$ gives no (complete) information about the answer. JSD
ranges from 0 to 1 with 0 (1) corresponding to complete (no) overlap between
$P$ and $Q$. Written in terms of Shannon entropy, $\text{JSD}(P||Q)=H(M)-
\frac{1}{2}(H(P)+H(Q))$.

\end{section}

\begin{section}{JSD Invariance for an Optimum Classifier}

In High Energy Physics (HEP) and elsewhere, one often performs
signal-background classification by choosing a supervised learning algorithm
(neural network, decision tree, support vector machine, etc.) in which one
first optimizes (trains) on the known samples. One is effectively using the
samples to try to recreate the pdfs that generated them. If one had those
underlying pdfs, no ML algorithm would be needed. Given the \varset\ for any
unknown event, one would simply query the pdfs to determine the relative
signal-background density for that position in variable space, which is also
the relative signal-background likelihood for the event. The closer the model
created by the ML algorithm is to the true pdfs, the more accurate it will be
in classifying future events. With signal and background treated as two simple
hypotheses, the Neyman-Pearson Lemma tells us that the likelihood ratio for
the event's point in variable space is the uniformly most powerful test for
its type~\cite{Neyman}.

Although in the studies we report below we used neural networks as the ML
algorithm, the results on separability are independent of this choice. We
chose them because they are widely used in HEP and in other fields, and
because excellent implementations are readily available. We used the
Tensorflow implementation with Keras as a front end~\cite{TensorFlow,Keras}.
The networks all had an input layer with a number of nodes equal to the number
of input variables ($n$), two hidden layers with varying numbers of nodes, and
an output layer with a single node. This commonly used architecture results in
a network that reduces the $n$-dimensional input space to a 1-dimensional
output space. For a fully optimized network, the output value for an event
will match the relative signal-background density for that event's location in
the original $n$-dimensional space~\cite{Richard, Zhang}. If one views the
input and output spaces as binned, a fully optimized network effectively
collects the bins from the input space that have equal signal-background
ratios and merges them into a single bin in the output space. That this
transformation will leave JSD unchanged is easy to see qualitatively. Recall
that JSD between the signal and background can be expressed in terms of MI as
the reduction in uncertainty about the event's type that comes from knowing
the \varset, or equivalently, the event's bin in variable space. Because that
reduction in uncertainty depends only on the signal-background ratio of the
bin, combining bins with the same signal-background ratio does not change MI.
We show this more formally in the appendix.

\end{section}

\begin{section}{Applying JSD and MI to measure separability and separation}

An optimum algorithm will leave JSD invariant as it transforms the signal and
background distributions from $n$-dimensional input to 1-dimensional output.
Both JSD between the input distributions (\jsdbefore\kern-0.7ex) and JSD
between the output distributions (\jsdafter\kern-0.7ex) will be useful.
\jsdbefore will define how much discrimination we can ever achieve with
\varset, and so lets us compare different potential sets.
\jsdafter\kern-0.7ex, when compared to \jsdbefore\kern-0.7ex, will tell us
whether or not an algorithm has extracted all possible information from the
input variables it was given.

It is useful to be precise about JSD as a figure of merit (FOM). The values
for \varset\ that maximizes \jsdbefore will reduce the information contained
in the signal-background answer more than the values from any other set of
variables. If \jsdafter=\jsdbefore\kern-0.7ex, then the output value of the
algorithm reduces the information in the answer as much as the input variables
do. Maximizing this FOM does not guarantee an optimum for other FOM's, such as
$S^2/B$ (where $S$ and $B$ are signal and background efficiency), or the area
under an accept--reject curve, or false-signal errors or false-background
errors. The optimum for any of these FOM's will generally not be the optimum
for the others, and so they cannot be optimized simultaneously. Thus, careful
choice of the cost function for an algorithm may still increase a particular
FOM. In practice, this is often a small effect, and the set \varset\ that
maximizes one of them is likely able to maximize the others. Further, the
requirement that \jsdafter=\jsdbefore will still guarantee that all available
information is being used.

Because the output signal and background distributions are 1-dimensional, it
is straightforward to calculate \jsdafter using $\text{JSD}(P||Q)=H(M)-
\frac{1}{2}(H(P)+H(Q))$, where $P$ and $Q$ are the signal and background
distributions, respectively. Because the input data are n-dimensional,
calculating \jsdbefore is far more difficult. Its equivalence to MI, however,
allows us to use recent advances for computing MI non-parametrically by
Kraskov, St{\"o}gbauer, and Grassberger (KSG)~\cite{Kraskov}. Unlike kernel
density approaches, KSG uses neighbor distances to estimate local density and
so avoids the intermediate step of finding a pdf for the variable space. They
point out that although nearest-neighbor methods have long been effective for
making non-parametric estimates of entropy, one cannot calculate MI from those
estimates by simply using $\text{JSD}(P||Q)=H(M)- \frac{1}{2}(H(P)+H(Q))$
because the errors on the individual entropies will not cancel. They developed
a dedicated nearest neighbor method for computing MI.

The well-known estimator for relative entropy by Kozachenko-Leonenko~\cite{KL}
is given by $$ S \approx \frac{d}{M}\sum_i\log(\lambda_i) - \psi(k) + \psi(N)
+ \log(V_d)$$ where d, N are the dimension and number of points in the sample
space, $V_d$ is the volume of the unit ball of dimension $d$, $M$ is the
number of mean non-vanishing distances $\lambda_i$ of nearest-neighbors $k$,
and $\psi(k)$ is the digamma function $$\psi(x) =
\frac{\Gamma'(x)}{\Gamma(x)}$$ where $\Gamma(x)$ is the gamma function. When
$x$ is an integer we can write this as $$\psi(n) = \sum_k^{n-1}\frac{1}{k} -
\gamma$$ where $\gamma$ is the Euler-Mascheroni constant; $\gamma \approx
0.57721$. It has been studied extensively and works well even in high
dimensions~\cite{Delattre}. To use it for MI however, one would need to find
the differences between several entropies. KSG point out that the errors on
these entropies will not necessarily cancel, and can lead to unstable or
non-physical MI values. In the spirit of KL's relative entropy estimator, KSG,
developed an MI estimator given by $$ \text{MI}(x;y) \approx \psi(k) + \psi(N)
- \frac{1}{k} - \langle \psi(n_x) + \psi(n_y)\rangle $$ where the $\psi(k)$
are still the digamma function except now one averages over the digamma
functions $\psi(n_x)$ and $\psi(n_y)$, where $n_x$, $n_y$ are the number of
nearest neighbors in the marginal spaces of $x$ and $y$ that are within the
mean $k$-nearest neighbor distance in the joint space. Thus, one first
constructs the joint space of $x$ and $y$ and then for each point finds the
nearest-neighbor distance. Then one averages over all of the digamma
evaluations for the numbers of nearest-neighbors in the marginal spaces ($x$
and $y$) that fall within those distances.

For this calculation, we use the NPEET software package (Non-Parametric
Entropy Estimation Toolbox)~\cite{VerSteeg,NPEET}, which includes a python
implementation~\cite{Python,Numpy} of the KSG method. We made one modification
to its implementation. Typically MI is computed among sets of continuous
variables, but here we compute it between the set \varset\ and a binary
answer. The problem is that when performing the neighbor count in the marginal
space of the answer, the points all have values of either $+1$ (signal) or
$-1$ (background). To count the neighbors, NPEET uses a KD-Tree approach, but
because it is unable to find reasonable bifurcation values, it reverts to
brute force and becomes very slow, even if a small amount of noise is added to
the values, as suggested by KSG.\@ However, since the marginal space only
contains the answer $\theta$, we already know the number of neighbors it
should find and so we modified the routine to use that number directly. The
result is an algorithm that finds \jsdbefore very quickly. For the slowest of
the studies described in later sections, we found stable values of \jsdbefore
in $\sim30$ seconds. This is a much faster way to measure the merit of a set
\varset\ than fully optimizing an ML algorithm then calculating an FOM on its
output. Table~\ref{tab:jsd_timing} shows the time needed to calculate
\jsdbefore for a range of variables and numbers of events using the Higgs
simulated data sample. (Details on the sample will be discussed later.) The
\jsdbefore values are also very stable, with a relative RMS variation of 0.4\%
over ten independent subsamples.

\begin{table}[!htb]
  \centering
  \begin{tabular}{|c|r|c|}
	\hline
	\# Dimensions & \# Points & \jsdbefore timing [sec]\\ \hline
	1  &	  100,000 & $\ph\ph4$ \\
	1  &	1,000,000 & $\ph72$ \\
	2  &	  100,000 & $\ph\ph5$ \\
	2  &	1,000,000 & $\ph74$ \\
	5  &	  100,000 & $\ph\ph6$ \\
	5  &	1,000,000 & $\ph97$ \\
	10 &	  100,000 & $\ph14$ \\
	10 &	1,000,000 & $311$ \\
	\hline
  \end{tabular}

  \caption{Times needed to compute \jsdbefore for various numbers of events
  and variables for the Higgs sample. \jsdbefore calculation times for the
  slowest tests described in later sections, took $\sim30$ seconds, as stable
  values could be found with far fewer than 1,000,000 events. Calculations
  were run on a fairly generic desktop computer with a fourth-generation Intel
  i5~cpu and 8~Gb RAM. \label{tab:jsd_timing}}

\end{table}

Though not directly addressed in KSG, we were concerned that variables with
widely different ranges of values might effectively be given different weight
in the MI calculation. If, for example, one has a much larger scale than the
others, its values will be more important when finding the distance to the
nearest neighbor. Therefore, before computing MI, we compute for a random
sample of events the RMS average distance to its nearest neighbor in each
dimension. We then scale each dimension of the data to force those values to
match. Note that this scaling is used only to calculate \jsdbefore and is
independent of the scaling one typically applies to neural network input
variables.

MI has been studied extensively for feature selection. For example, the
``Mutual Information Feature Selection'' (MIFS) algorithm by
Battiti~\cite{Battiti} is widely used to find non-redundant variables using a
greedy algorithm and pairwise computation of MI between variables. While often
effective, MIFS has limitations, as pairwise variable comparisons may ignore
important correlations among larger groups of variables. Further, until recent
improvement in calculating MI, errors in entropy calculations could reduce its
effectiveness. Work since Battiti has further improved feature selection
algorithms by, for example, more efficiently choosing which MI values should
be computed~\cite{Kwak}--\cite{Fleuret}. Other recent work on feature
selection has taken advantage of improvements in calculating MI in higher
dimensions to directly study larger subsets of variables, and to study the MI
between the variables and the class answers~\cite{Bonev_b}--\cite{Chow}, as we
do here. In this work, by modifying the MI calculation, and by showing the
invariance of MI under an optimum algorithm, we are able to study its
importance as a practical limit of separability. In the next section, we apply
this in several cases.

\end{section}

\begin{section}{Tests}

Our procedure for each of the tests below is:
\begin{enumerate}
\item For the input data, compute MI between the answer and the
n-dimensional mixed signal-background samples (\jsdbefore\kern-0.7ex)
\item Optimize the classification algorithm on the samples.
\item Compute JSD between the output signal and background distributions
produced by the algorithm (\jsdafter\kern-0.7ex)
\end{enumerate}

\begin{subsection}{Verifying JSD Invariance}

We first tested the claim that JSD separation between signal and background is
unchanged by processing the events through a classification algorithm. For
this test we generated Gaussian distributions in five dimensions
($P(\vec{x})=\frac{1}{{\sqrt {2\pi\sigma^2 } }}e^{-\frac{{d^2}}{2\sigma^2}}$,
where $d=\sqrt{\sum_{i=1}^{5} (x_i - \mu_i)^2}$) for both signal and
background, and varied the separation between them. The 25 signal (background)
distributions each had 1000 events, $\sigma=1.0$, and $\mu_i$ that varied
between 0.04 and 1.0 (-0.04 to -1.0) in 25 steps. Combining the samples
resulted in 25 signal-background samples each with 2000 events and a distance
between the signal and background means ($\Delta\mu$) that varied from 0.08 in
sample 1 to 2.0 in sample 25.

For each sample, we used 1400 events as a training set to optimize a
fully-connected feedforward neural network with backpropagation learning. The
network had an input layer with 5 nodes, two hidden layers with 11 and 4 nodes
respectively, and an output layer with one node. We trained the network for 10
epochs then used the remaining 600 events as a test set to verify that it had
not been overtrained.

The critical test is whether or not JSD, as a measure of signal-background
separation, remains invariant as the neural-network transforms the data from
five dimensions to one dimension. For each sample we first calculated JSD for
the 1400 training events in the five-dimensional variable space
(\jsdbefore\kern-0.7ex). The mean values and errors from ten independent
subsamples are shown as a dashed line in Figure~\ref{fig:gauss5d_sliding}. As
expected, \jsdbefore increases with increasing signal-background separation.
For each sample we next calculated the JSD of the network output
(\jsdafter\kern-0.7ex) using the 600 testing events. Using the training sample
to calculate \jsdbefore and the testing sample to calculate
\jsdafter\kern-0.7ex, ensures that the two sets are statistically independent.
The results for \jsdafter are shown as a solid line in
Figure~\ref{fig:gauss5d_sliding}, where again the results are the means from
ten independent subsamples. As expected, the neural-network is better able to
separate the samples with larger $\Delta\mu$. The key result is the excellent
agreement between \jsdbefore and \jsdafter across the entire range of
$\Delta\mu$, from nearly complete overlap to nearly complete separation. In
every case, the final separation that the neural network achieves
(\jsdafter\kern-0.7ex) is as good, but never better, than the predicted
separation from looking only at the raw data in five dimensions
(\jsdbefore\kern-0.7ex). We also tested cases where we prevented the network
from reaching optimum separation by either using too few nodes or by not
allowing enough training cycles. In every case we found,
\jsdafter$<$\jsdbefore\kern-0.7ex.

The results for JSD fluctuate very little between subsamples. Here, and in the
studies discussed below, we find that in independent subsamples, the RMS
variation for \jsdbefore typically corresponds to $<1\%$ and is occasionally
as high as $\sim4\%$ relative uncertainty. The variation in \jsdafter is
typically $3$--$5\%$; its variations are dominated by variations in how each
of the neural networks converges.

\begin{figure}[!htb]
\rotatebox{0}{\includegraphics[width=3.4in]{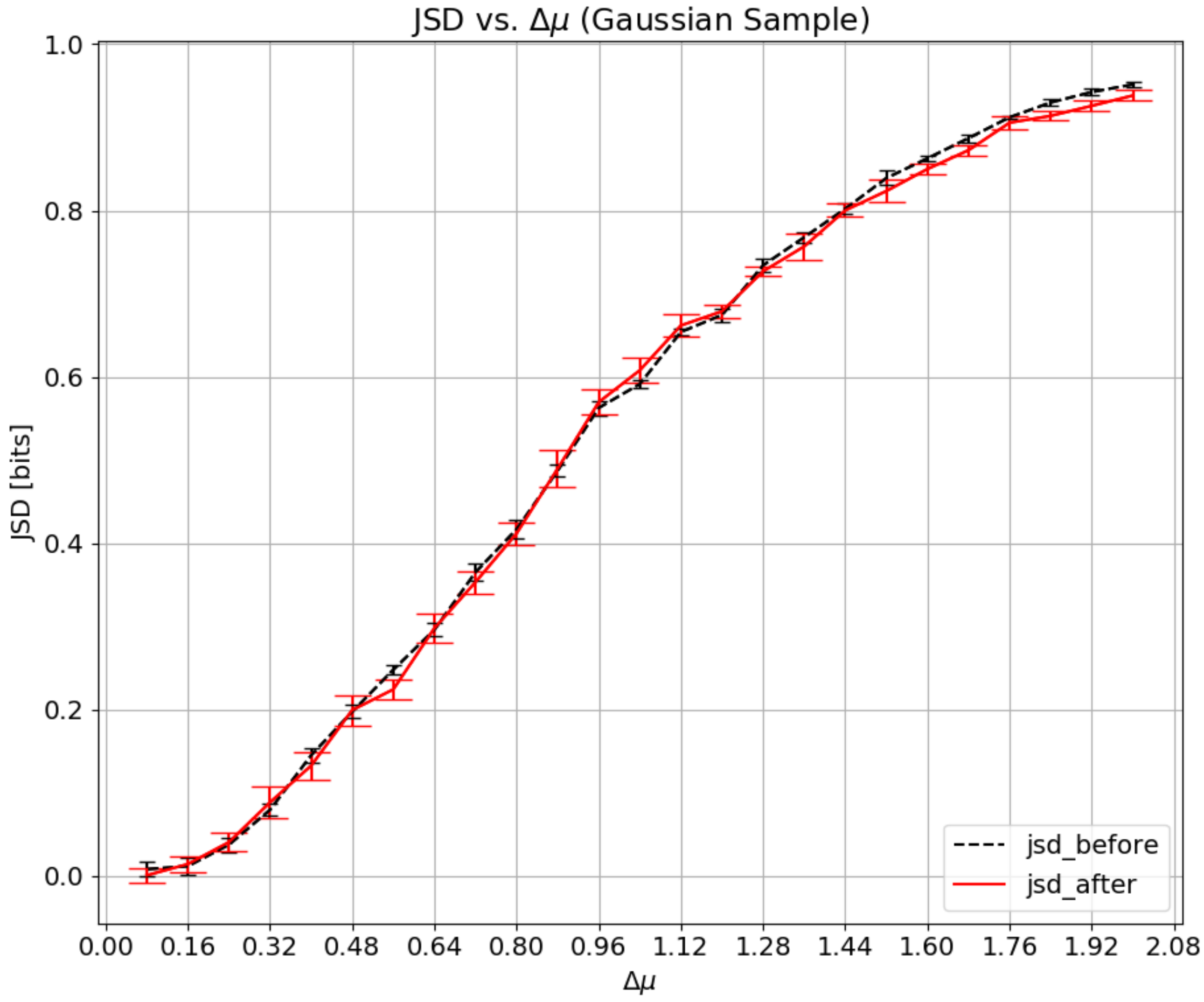}}

\hfill \caption{Comparison of \jsdbefore (black, dashed) to \jsdafter (red,
solid) for Gaussian distributions in 5-dimensions in which the signal and
background separation ranges from nearly complete overlap (small $\Delta\mu$)
to nearly complete separation (large $\Delta\mu$). The achieved separation by
the ML algorithm (\jsdafter\kern-0.7ex) tracks the predicted separation
(\jsdbefore\kern-0.7ex) over the entire range. The results are the means from
ten independent samples.\label{fig:gauss5d_sliding}}

\end{figure}

\end{subsection}

\begin{subsection}{Verifying JSD Increases only When Additional Variables are
Non-Redundant}

Here we tested the claim that JSD is a fundamental measure of separability and
hence should increase with additional discriminating variables only if they
add information not already available in other variables. We generated
Gaussian distributions from the same parent as in the previous study. Each
signal (background) sample had 5,000 events, a mean in each dimension of 1.0
(-1.0) and a variance in each dimension of 1.0. As in the previous test, we
compared \jsdbefore to \jsdafter\kern-0.7ex. Here however, we added the five
discriminating variables one at a time, then compared \jsdbefore to \jsdafter
following the addition of each one. Figure~\ref{fig:gauss5d} shows that adding
additional variables increases the predicted upper limit of separation and
also the separation that the neural network achieves. As in the previous test,
\jsdbefore and \jsdafter are consistent with one another (though there is some
evidence that with five variables the network did not completely reach the
optimum). As before, the mean values and errors are from ten independent
subsamples.

\begin{figure}[!htb]
\rotatebox{0}{\includegraphics[width=3.4in]{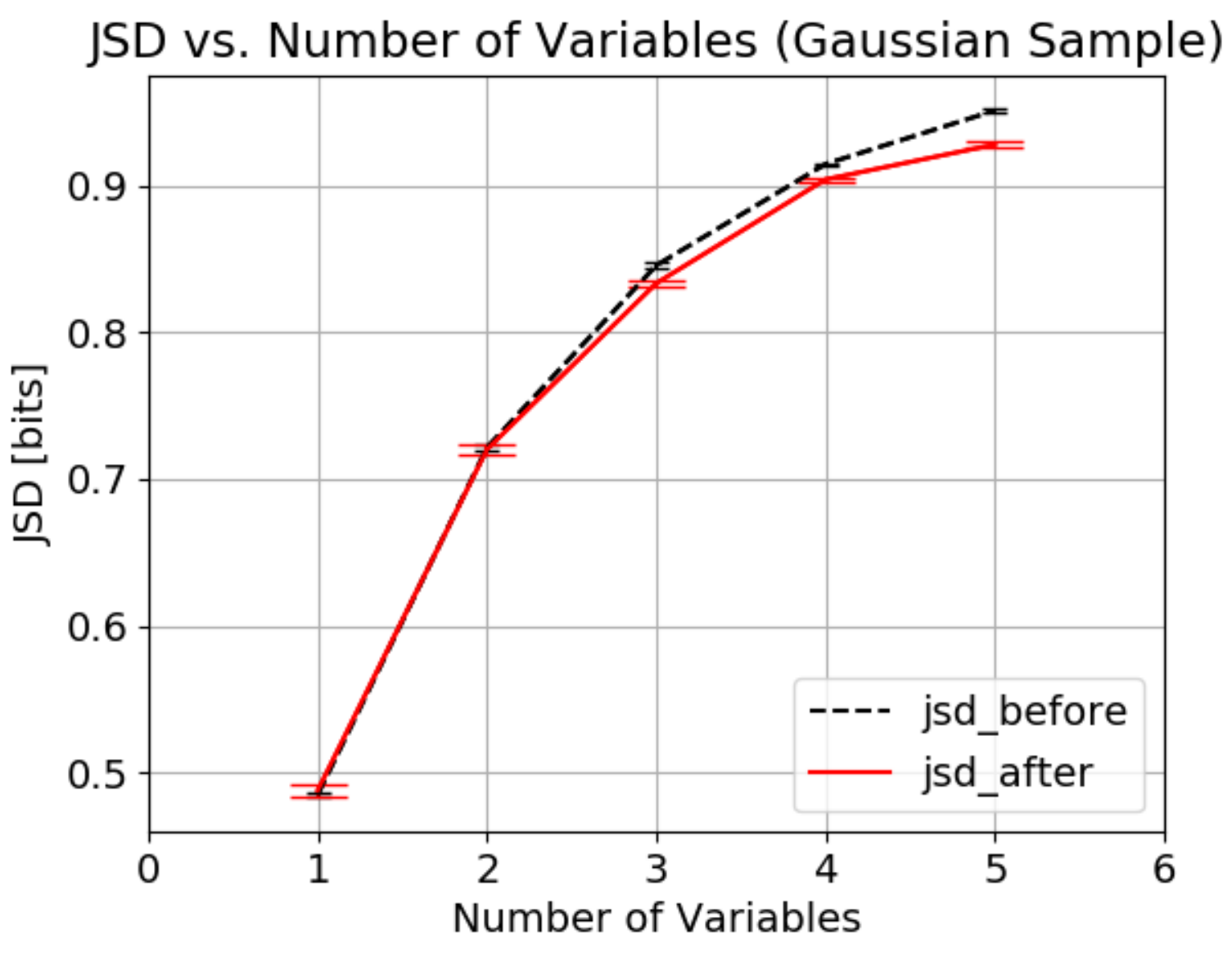}}

\hfill \caption{Comparison of \jsdbefore (black, dashed) to \jsdafter
(red, solid) for Gaussian signal and background distributions in
5-dimensions. The signal-background separation is fixed, and the five
discriminating variables are added to the ML algorithm one at a time. The
achieved separation by the ML algorithm (\jsdafter\kern-0.7ex) improves with
each additional variable, and it tracks the predicted separation
(\jsdbefore\kern-0.7ex). The results are the means from
ten independent samples.\label{fig:gauss5d}}

\end{figure}

To test our ability to identify redundant information, we repeated the above
test, except that after adding the first three variables, one at a time, we
added two variables that were made to be functions of the first three. Then
finally we added the last two independent variables. Thus the fourth and fifth
added variables contained no new information.
Figure~\ref{fig:gauss5d_redundant} shows that JSD improves as one adds
information from variables 1 through 3, but then does not improve further when
adding the next two. The last two variables bring JSD back up to the level it
reached in the previous test. As before, \jsdbefore and \jsdafter track each
other well throughout. This ability to identify the underlying total
information content available for distinguishing between the classes even in
cases where variables may be functions of one another is important for feature
selection. Knowing \emph{a priori} the maximum possible separability allows
one to judge how much information, if any, is lost when choosing a subset of
all available variables for class separation. One does not need to know how
variables may depend on one other to determine the maximum achievable
separation between event types.

\begin{figure}[!htb]
\rotatebox{0}{\includegraphics[width=3.4in]{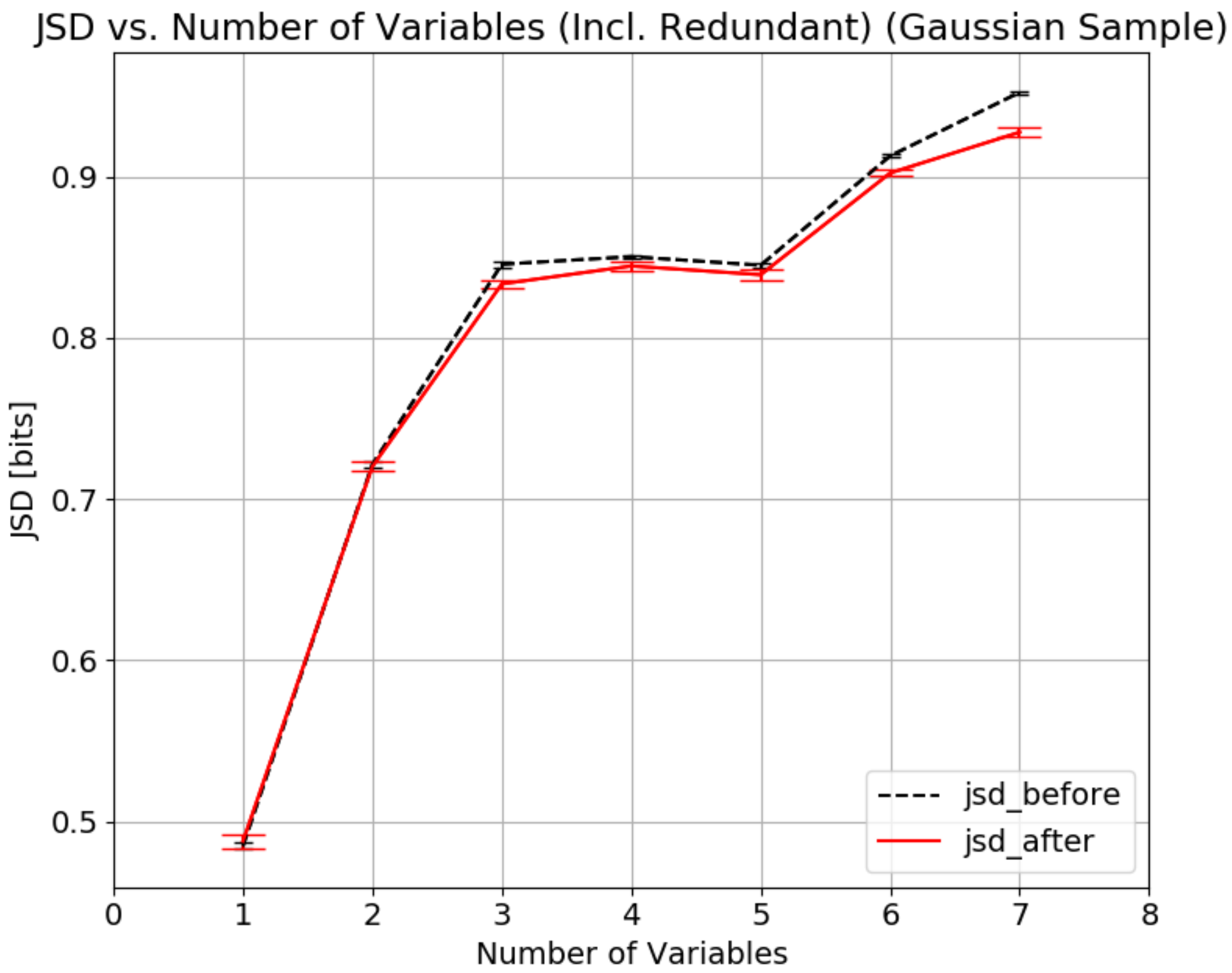}}

\hfill \caption{This figure shows the same comparison made in
Figure~\ref{fig:gauss5d}, except that here, variables four and five are
functions of the first three. \jsdbefore correctly indicates that those two
variables do not increase the separability of the samples. Adding the last two
independent variables (now, the sixth and seventh) brings the predicted
separability back to the level seen in Figure~\ref{fig:gauss5d}. The results
are the means from ten independent samples.\label{fig:gauss5d_redundant}}

\end{figure}

\end{subsection}

\begin{subsection}{Simulated Data for Particle Physics Searches}

To test our approach on more realistic data, we used two simulated datasets
produced by Baldi, Sadowski, and Whiteson (BSW) for their study on particle
physics search methods~\cite{BSW}. They produced two simulated (Monte Carlo)
datasets that mimic information that would be available in a particle physics
search at the Large Hadron Collider. Madgraph5~\cite{madgraph} was used as the
generator and Pythia~\cite{pythia} was used for hadronization and showering. The
detector response was then simulated using DELPHES~\cite{delphes}.

One set mimics a Higgs search, and the other a Supersymmetry (SUSY) search.
Both sets have kinematic variables for discriminating signal from background
that are typical of those available in a Large Hadron Collider (LHC) dataset;
both datasets are publicly available~\cite{mlrepo}. The data have
``low-level'' and ``high-level'' variables. The low-level variables are
kinematic event features expected to be helpful in distinguishing signal from
background. The high-level variables are functions of the low-level variables,
hence they contain no new information and so in principle are extraneous. In
practice, however, classifiers are often unable to extract all available
information from low-level variables, and so derived quantities often improve
separation performance. BSW investigated the effects of a neural-network's
architecture and learning methods on its signal-background separation
performance, and also on its ability to use low-level variables without
reliance on high-level variables. Below, we find that our predicted
separability limits for their datasets are just above the maximum separations
that our networks were ultimately able to reach. In the case of the Higgs
sample, we were also able to compare our limits to the separation their
network achieved, and again find that they are consistent. We also find that
including or excluding the high-level variables has no effect on our limit,
verifying the important principle that it is a limit on fundamental
signal-background separability, unaffected by redundant information.

\begin{subsubsection}{Higgs Sample}

For signal events in the Higgs sample, a theoretical neutral Higgs boson is
produced through the fusion of two gluons. The neutral Higgs decays into a
charged Higgs and a $W$ boson. The charged Higgs then decays into a $W$ and
the Standard-Model (SM) Higgs. This SM Higgs then decays predominantly into
charged $b$ quarks. The products are thus a pair of charged $W$'s and a pair
of charged $b$ quarks which further decay or hadronize respectively. The
process for the background events also produces a pair of charged W's and a
pair of b quarks, but without the intermediate Higgs state. This leads to
kinematic differences that can be used for signal-background discrimination.
The events are described by 21 low-level and 7 high-level variables. Because
several of the low-level variables are discrete, and the Kraskov estimator for
mutual information is designed for continuous variables, we decided for this
study to make comparisons only with the high-level variables. (We expect that
it will be straightforward to modify the algorithms to work with both
continuous and discrete input.) Because the difference between signal and
background events is the presence or absence of intermediate Higgs particles,
all seven of the high-level variables are mass estimates made by combining the
individual observed particles from the $W$ decays and the groups of particles,
known as jets, produced by the $b$ quarks.

We divided a 5-million-event sample into 10 equal subsets. As in the previous
test, we tracked the separability limit as we added high-level variables one
at a time. After each was added, we optimized a neural network and measured
the separation it achieved. The networks, had an input layer with a number of
nodes equal to the number of input variables, two hidden layers with 34 and 27
nodes, and an output layer with one node. We trained the networks for 1000
epochs using, 70\% of the sample for training and 30\% for testing. Our
optimized networks approached but never surpassed the predicted separability
limit (see Figure~\ref{fig:higgs_jsd}).

\begin{figure}[!htb]
\rotatebox{0}{\includegraphics[width=3.4in]{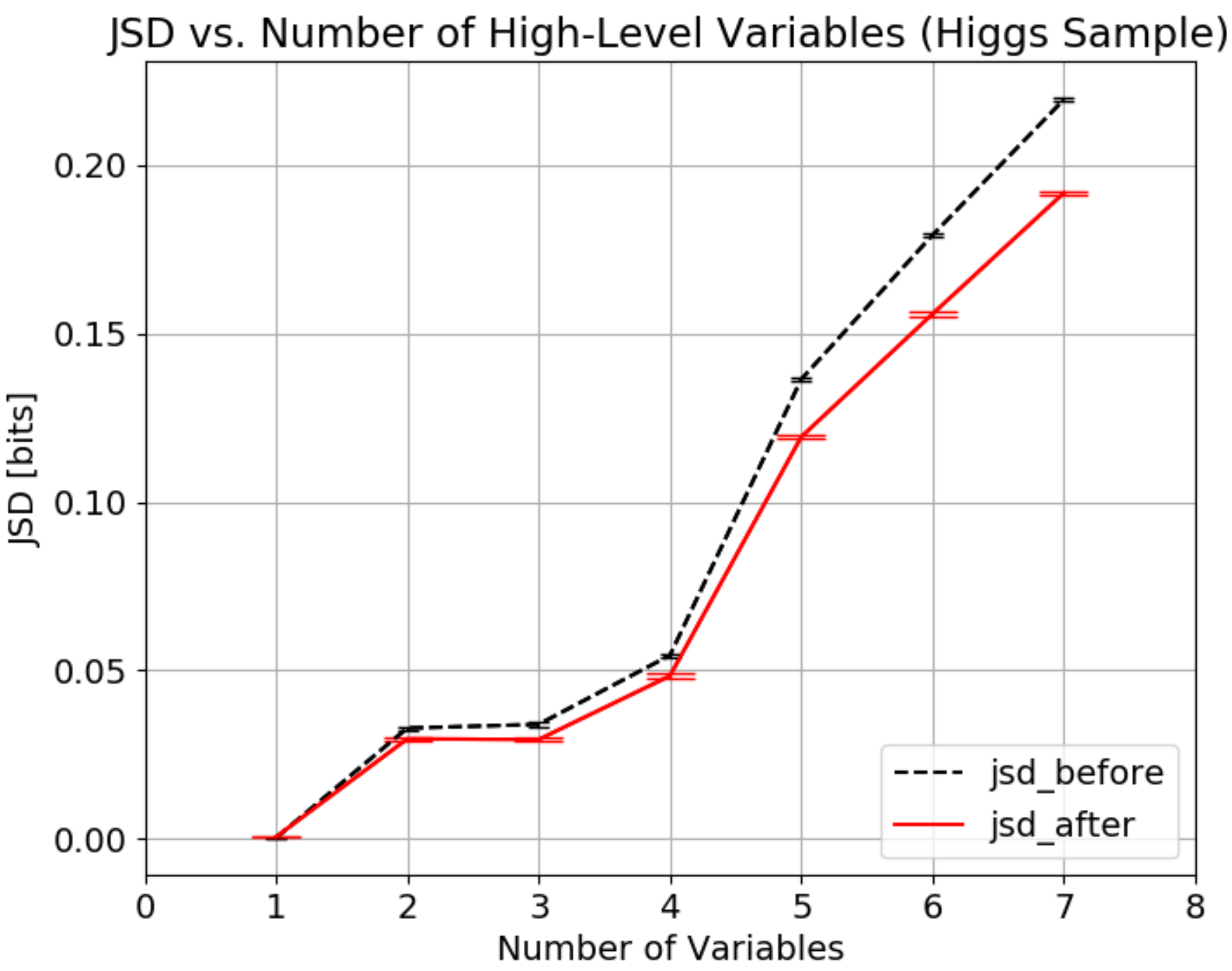}}

\hfill \caption{Comparison of \jsdbefore (black, dashed) to \jsdafter (red,
solid) for the Higgs sample as high-level discriminating variables are added
one at a time. \jsdbefore increases as each of the seven high-level variables
is added. The achieved separation by the ML algorithm (\jsdafter\kern-0.7ex)
tracks the predicted separation (\jsdbefore\kern-0.7ex) reasonably throughout,
without ever exceeding the predicted separations. The results are the means
from ten independent samples.\label{fig:higgs_jsd}}

\end{figure}

We also used the performance of the networks in BSW as a benchmark. BSW
converted the output distributions of their networks into background rejection
vs.\ signal efficiency curves, commonly known as Receiver Operating
Characteristic (ROC) curves. They then used the total area under the ROC curve
(AUC) as their figure of merit. Larger values of AUC correspond to better
signal-background separation. We compared the AUC values from our networks to
those in BSW to ensure that we were using their data correctly, and also to
verify that our networks were reasonably well optimized, as they took
extensive efforts to verify that their networks had converged fully. As
expected, in the cases where their networks were significantly larger, they
reached slightly higher AUC values (see Table~\ref{tab:higgs_jsd}).

We next attempted to compare their achieved separation to our separability
limit. To do this, we used the bin-values from their ROC curves to reconstruct
network output distributions, which then allowed us to compute the \jsdafter
FOM for their networks\footnote{We thank the BSW authors for providing, where
possible, the datapoints of their ROC curves}. They first tested a small
network that they noted did not achieve a high-level of discrimination. We
estimate that its output corresponds to a separation of
\jsdafter\kern-0.7ex=0.15~\footnote{For this network, we did not have the
datapoints for their ROC curve, and so digitized their plot.}, which is less
than our predicted limit of \jsdbefore\kern-0.7ex=0.22. They then used a much
larger network and we found that its output corresponds to a separation of
\jsdafter\kern-0.7ex=0.22. The key point is that their large well-optimized network
reaches but does not surpass the separability limit that we predicted from the
raw data. The full results are shown in Table~\ref{tab:higgs_jsd}.

\begin{table}[!htb] \centering 
	\begin{tabular}{|l|c|} \hline & High-level vars only\\ \hline 
		\jsdbefore (this work) & 0.22 (0.001) \\ \hline 
		\jsdafter (this work) & 0.19 (0.002) \\ 
		\jsdafter (BSW, shallow network) & 0.15 \\ 
		\jsdafter (BSW, deep network) & 0.22 \\ \hline \hline 
		AUC (this work) & 0.78 \\ 
		AUC (BSW, shallow network) & 0.78 \\ 
		AUC (BSW, deep network) & 0.80 \\
\hline \end{tabular}

  \caption{Comparison of JSD and AUC values for the 7 high-level variables in
  the Higgs sample. Lines two and three show that neither of the two smaller
  networks (ours or the shallow network in BSW) reach the predicted separation
  (line one). The larger network in BSW (line four) reaches but does not
  surpass the prediction. The last three lines show AUC values that we used to
  verify our networks against those in BSW. Lines one, two, and five are the
  means from ten independent samples and the values in parentheses are the RMS
  variations among them. \label{tab:higgs_jsd}}

\end{table}

\begin{subsubsection}{Supersymmetric (SUSY) Sample}

For the SUSY sample, the signal events are a process in which supersymmetric
$\chi^\pm$ particles are produced and then decay to $W$ bosons and a
supersymmetric $\chi^0$. The $W$'s subsequently decay to charged leptons and
neutrinos. The $\chi^0$ and the neutrinos are not directly observable and
their presence is inferred from momentum imbalance and missing energy in the
detector. The background events are from a Standard Model process in which two
$W$ bosons are produced and each decays to a charged lepton and a neutrino.
The signal and background events can be distinguished because they differ in
the number of invisible particles and in the kinematics of the decays. The
events are described by eight low-level variables: For each of the two
leptons, its angle is described by two variables and its momentum transverse
to the beamline by one variable. Two additional low-level variables give the
energy and momentum imbalance caused by the undetected particles. There are
ten high-level variables derived from these eight. The details of these are
unimportant for our studies because, as functions of the eight low-level
variables, they contain no additional information.

We divided the 5-million-event sample into 10 equal subsets and optimized
neural networks on the low-level variables, the high-level variables, and both
low-level and high-level combined (all). The networks had two hidden layers
with 24 nodes and 17 nodes respectively and a single output node. As required,
we adjusted the number of input nodes to match the number of input variables.
We trained the networks for 100 epochs using, as before, 70\% of the sample
for training and 30\% for testing.

As in the Higgs study, to ensure that our networks were well optimized, we
compared the AUC values to those in BSW~\footnote{We were unable to
reconstruct a \jsdafter value for their networks, as we did in the Higgs
study, because their publicly available SUSY dataset does not have all the
variables that are used in their networks.}. Figure~\ref{fig:susy_auc} shows
the AUC performance of the networks on low-only, high-only, and all variables.
Table~\ref{tab:susy_jsd} compares these AUC values to the networks from BSW.

\begin{figure}[!htb]
\rotatebox{0}{\includegraphics[width=3.4in]{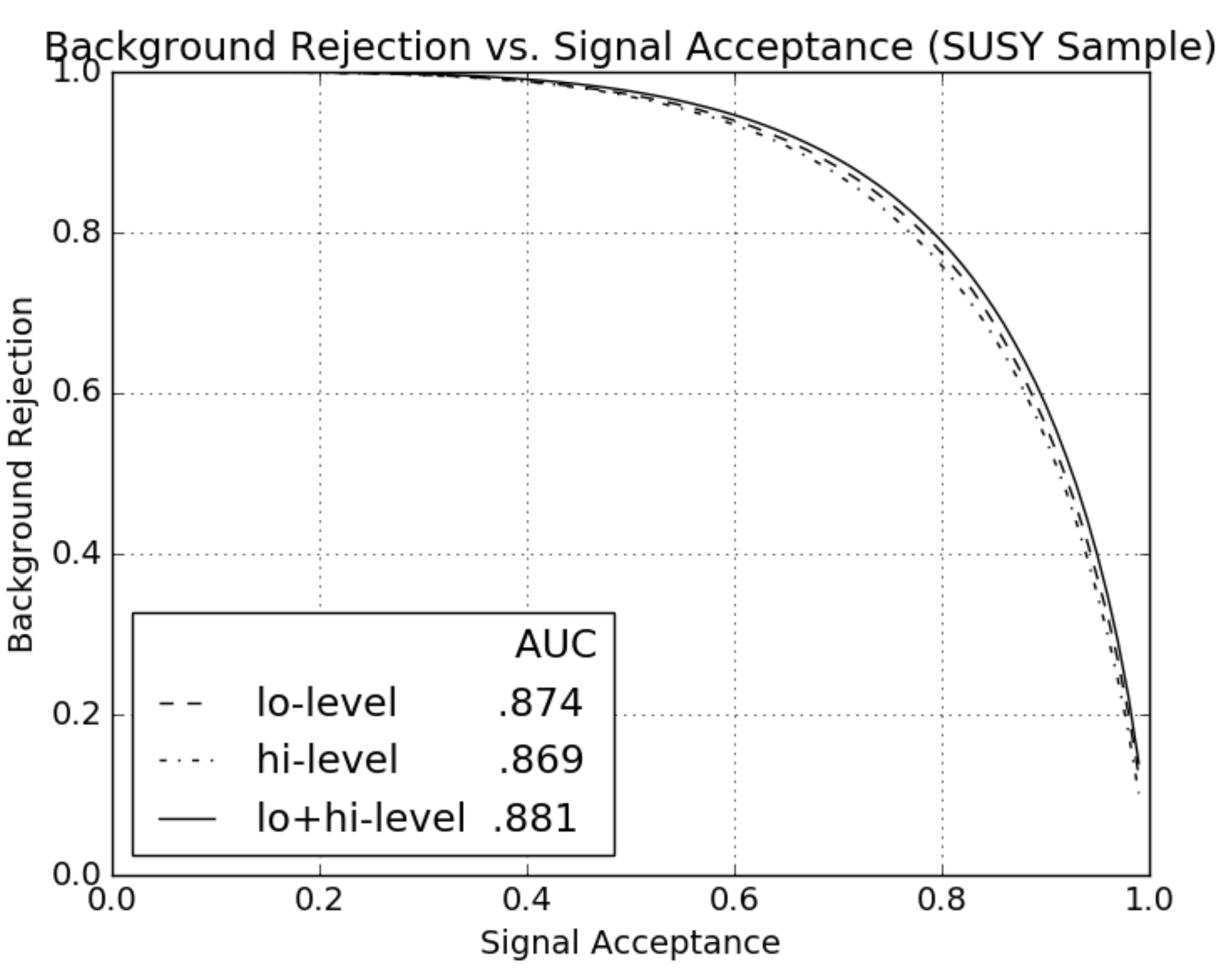}}

\hfill \caption{Neural network background rejection vs.\ signal acceptance ROC
curves for the SUSY sample. The low-level only, high-level only, and
low+high-level sets are shown as dashed, dot-dashed, and solid curves
respectively. The ROC curves from our networks were used to verify the
reasonableness of our networks against those in BSW. The AUC results from
these curves also appear in Table~\ref{tab:susy_jsd} \label{fig:susy_auc}}

\end{figure}

We next made the key comparison of \jsdbefore to \jsdafter\kern-0.7ex.
Table~\ref{tab:susy_jsd} shows the results for networks trained on the 8
low-level variables, the 10 high-level variables, and all 18 variables. For
all three cases \jsdafter approaches, but never surpasses
\jsdbefore\kern-0.7ex.

\begin{table}[!htb]
  \centering
  \begin{tabular}{|l|p{13mm}|p{14mm}|p{11mm}|}
	\hline
	& Low only & High only & Both \\ \hline
	\jsdbefore (this work) & 0.36 \newline (0.002) & 0.36 \newline (0.002) & 0.37 \newline (0.002)\\ \hline
	\jsdafter (this work) & 0.35 \newline  (0.004) & 0.35 \newline (0.005) & 0.36 \newline (0.005) \\
	\hline
	\hline
	AUC (this work) & 0.87 & 0.87 & 0.88 \\
	AUC (BSW, shallow network) & 0.86 & 0.86 & 0.88 \\
	AUC (BSW, deep network) & 0.88 & 0.87 & 0.88 \\
	\hline
  \end{tabular}

  \caption{Comparison of JSD and AUC values for the SUSY sample. The top line
  shows the \jsdbefore values for the 8 low-level, the 10 high-level, and the
  18 combined variable sets. Line two shows that our neural network separation
  reaches but does not surpass the predicted separability. The last three
  lines show AUC values that we used to verify our networks against those in
  BSW. Lines one through three are the means from ten independent samples and
  the values in parentheses are the RMS variations among
  them.\label{tab:susy_jsd}}

\end{table}

Figure~\ref{fig:susy_jsd} shows the results of a separate test on this dataset
in which we added variables one at a time, and compared \jsdbefore to
\jsdafter following the addition of each. As in the second test with
Gaussians, the predicted separation and achieved separation agree, and both
continue to improve as each additional low-level variable is included.
Importantly, \jsdbefore and \jsdafter both stop increasing once all the
low-level variables have been included. High-level variables (derived
quantities) do not change the fundamental separability of the datasets.

\begin{figure}[!htb]
\rotatebox{0}{\includegraphics[width=3.4in]{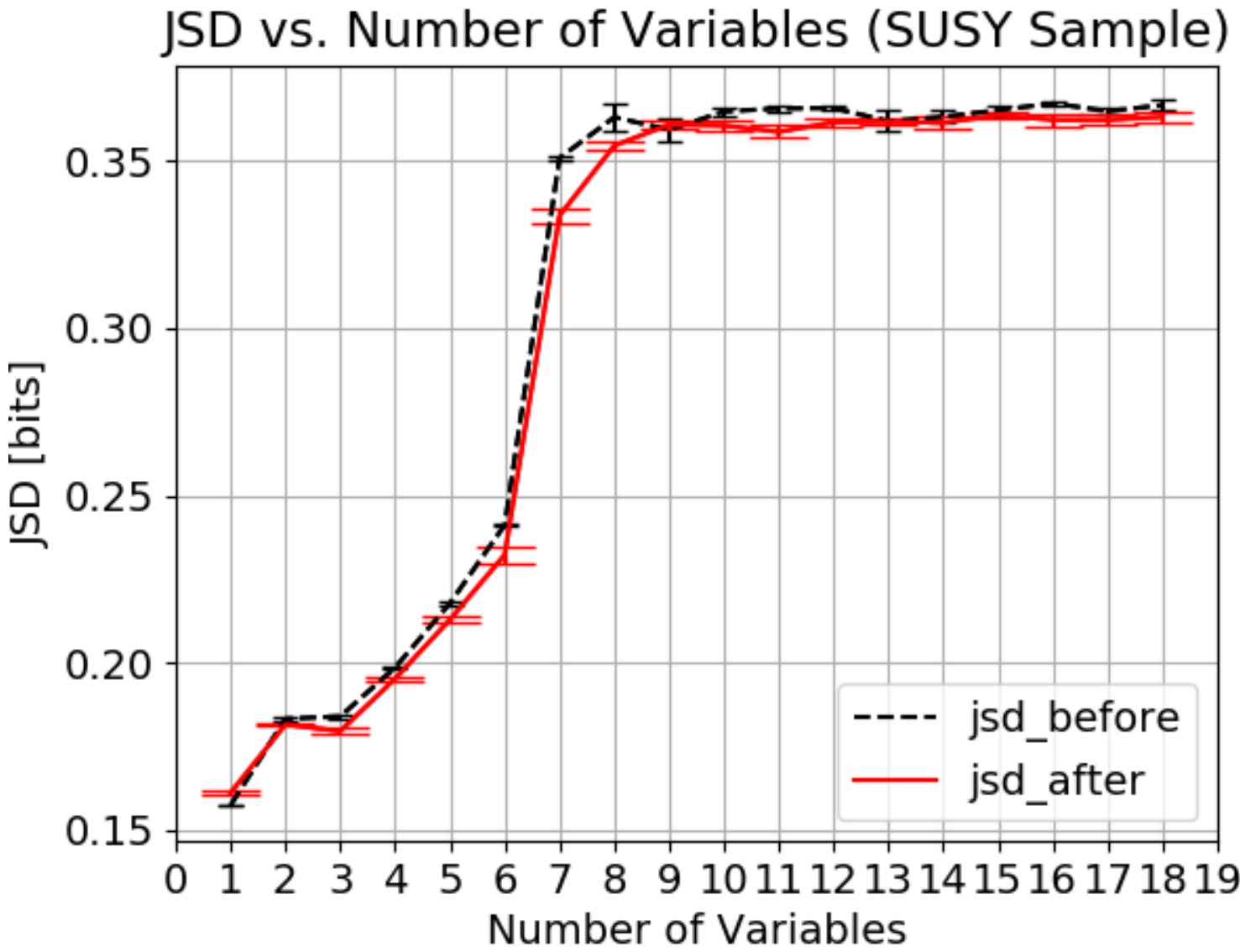}}

\hfill \caption{Comparison of \jsdbefore (black, dashed) to \jsdafter (red,
solid) for the SUSY sample as discriminating variables are added one at a
time. \jsdbefore increases as each of the eight low-level variables is added.
It does not increase further as the ten high-level variables are included.
This is expected because the high-level variables do not bring additional
discriminating information. The achieved separation by the ML algorithm
(\jsdafter\kern-0.7ex) tracks the predicted separation (\jsdbefore\kern-0.7ex)
throughout. The results are the means from ten independent
samples.\label{fig:susy_jsd}}

\end{figure}

\end{subsubsection}

\end{subsubsection}

\end{subsection}

\end{section}

\begin{section}{Conclusion}

Distinguishing signal from background based on a set of descriptive
quantities, is an important task in a wide range of fields. We have proposed
that using known samples, one can place a limit on the separability between
two event types by finding the Mutual Information between the known answer and
all discriminating variables. Equivalent to the Jensen-Shannon Divergence,
this limit is independent of the algorithm chosen to classify the events and
is only reachable if the algorithm preserves the relative signal-background
probability when transforming the data from the n-dimensional input space to
the 1-dimensional output space. We tested these limits on two datasets of
Gaussian distributions and then on two Monte Carlo samples generated to study
classification algorithms in particle physics searches.

This approach has substantial practical benefits for classification problems
both for feature selection and for algorithm evaluation. One benefit for
feature selection is that one can take a large number of potential variables
and quickly determine the separability that could be reached if all of them
were used at once. With that benchmark in hand, one can then monitor the
separability that would be lost by using any subset of the variables. Feature
selection also benefits because once a set is selected, one can quickly
investigate any potential new variables that might be added. The separability
limit will increase if the new variables bring new information but not if they
are effectively functions of those already being used. The benefits to
algorithm choice and optimization are clear. There are dozens of approaches to
classification tasks, and sub-methods for most of them. By comparing the
separation achieved by a given method to the separability limit calculated on
the input data, one knows whether or not the method has achieved the best
possible separation. In short, you know when you're done.

Our software for computing separability limits, along with documentation to
help users quickly calculate limits for their own datasets, is publicly
available at \url{https://github.com/albanyhep/JSDML}.

\end{section}

\begin{section}{Acknowledgments}

We thank Arial Caticha, Eric Dohner, Adam Fischer, Philip Goyal, Vivek Jain,
Kevin Knuth, Oleg Lunin, Udo von Toussant, Kevin Vanslette, Greg Ver Steeg,
and Daniel Whiteson for helpful discussions.

\end{section}

\begin{section}{References}

\end{section}

\begin{section}{Appendix}

\begin{subsection}{Proof of the Data Processing Inequality}

Starting from the discussion in Cover and Thomas~\cite{CoverThomas}, here we
give a concise proof of the data processing inequality in terms of Mutual
Information, which for $y=f(\{x\})$, and two categories of events
$\theta=\{+1,-1\}$, states that $\mathcal{M}[\theta;x] \geq
\mathcal{M}[\theta;y]$. This enforces the idea that a transformation
y=f(\{x\}) of the input variables can not increase their ability to
distinguish between the classes. Given variables $\theta$,$x$,$y$ which form a
Markov chain \begin{equation} \theta \rightarrow x \rightarrow y
\end{equation} implies the joint probability can be written \begin{equation}
p(\theta,x,y) = p(\theta)p(x|\theta)p(y|x) \end{equation} where $p(y|x)$ is
independent of $\theta$. We can appeal to the symmetric nature of the mutual
information to break it up in two different ways \begin{align}
\mathcal{M}[\theta;x,y] &= \mathcal{M}[\theta;x] +
\mathcal{M}[\theta;y|x]\nonumber\\ &= \mathcal{M}[\theta;y] +
\mathcal{M}[\theta;x|y] \end{align} but because of the Markov property that
$p(y|x,\theta) = p(y|x)$, we have that $\mathcal{M}[\theta;y|x] = 0$. And
since mutual information is always positive, $\mathcal{M}[\theta;x,y] \geq 0$
we conclude \begin{equation} \mathcal{M}[\theta;x] \geq \mathcal{M}[\theta;y]
\end{equation} with equality only when $\mathcal{M}[\theta;x|y] = 0$ or when
$y$ is a sufficient statistic for $x$ \begin{equation} p(\theta|x) =
p(\theta|y) \qquad \mathrm{i.e.} \qquad \theta \rightarrow y \rightarrow x
\end{equation}

\end{subsection}

\end{section}

\end{document}